\documentclass[useAMS,usenatbib,usegraphicx]{mn2e}
\usepackage{lscape}

\newcommand{\msun}{{{\rm M}_{\sun}}}

\title[Energetics of Cyg X-1]{Energetics of a black hole: constraints on the jet velocity and the nature of the X-ray emitting region in Cyg X-1}
\author[J. Malzac, R. Belmont and A. C. Fabian]{Julien Malzac$^{1}$\thanks{E-mail:malzac@cesr.fr}, Renaud Belmont$^{1}$ and Andrew C. Fabian$^{2}$\\
$^{1}$ CESR (Centre d'Etude Spatiale des Rayonnements),  Universit\'e de Toulouse [UPS], \\ CNRS [UMR 5187], 9 avenue du Colonel Roche, BP 44346, 31028 Toulouse Cedex 4, France\\
$^{2}$ Institute of Astronomy, Madingley Road, Cambridge, CB3 0HA}
\begin{document}

\date{Accepted 2009 August 16. Received 2009 August 7; in original form 2009 April 8}

\pagerange{\pageref{firstpage}--\pageref{lastpage}} \pubyear{2009}

\maketitle

\label{firstpage}

\begin{abstract}
We investigate the energetics of the jet and X-ray corona of Cyg X-1. We show that the current estimates of the jet power obtained from H${\alpha}$ and \mbox{[O\,{\sc iii}]} measurements of the optical nebula surrounding the X-ray source allow one to constrain the bulk velocity of the jet. It is definitely relativistic ($v>$0.1c) and most probably in the range  (0.3--0.8)c. The exact value of the velocity depends on the accretion efficiency. These constraints are obtained independently of,  and  are consistent with, previous estimates of the jet bulk velocity based on radio measurements. 
We then show that the X-ray emission does not originate in the jet. 
Indeed, the energy budget does not allow the corona to be  ejected to infinity at relativistic speed.  Rather, either a small fraction of the corona escapes to infinity, or the ejection velocity of the corona is vanishingly low. Although the corona could constitute the jet launching region, it cannot be identified with the jet itself. We discuss the consequences for various X-ray emission models. 
\end{abstract}


\section{Introduction}\label{sec:eandm}
\begin{figure}

\includegraphics[width=\columnwidth]{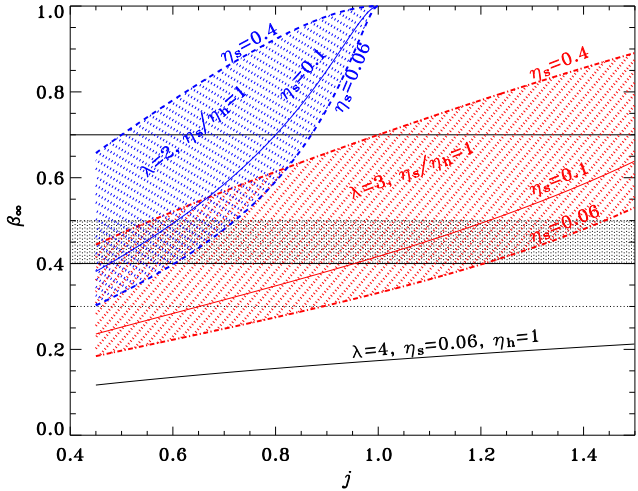}
\caption{Constraints on the jet terminal velocity as a function of the measured jet power. 
The conservative lower limit on $\beta_{\infty}$ as a function of $j$ is given by the black thick solid curve (assuming the maximum possible value for $\lambda=4$, and minimum value for the accretion efficiency in the soft state $\eta_{\rm s}=0.06$ and an efficiency of 1 in the hard state). 
The other curves show the dependence of $\beta_{\infty}$ on $j$ in the more reasonable case in which the efficiency remains constant across the state transition. 
The red hatched region shows the range of $\beta_{\infty}$ allowed for $\lambda=3$ and the accretion efficiency $\eta_{\rm s}$ ranging between 0.06 and 0.4. The blue hatched region shows the same for $\lambda=2$ . The full lines show the dependence of $\beta_{\infty}$ on $j$ for the fiducial case  $\eta_{\rm s}=0.1$. The {black thin solid lines} show the upper and lower limits provided by the radio imaging and radio X-ray correlations (Gleissner et al. 2004). The {  black thin dotted lines} show the constraints from the modelling of the super-orbital periodicity (Ibragimov et al. 2007). The horizontal grey stripe shows the overlapping region between those two constraints.}
\label{fig:betainfdej}
\end{figure}

Cyg X-1 is the prototype of black hole binaries. 
It is a persistent X-ray source, powered by accretion onto a black hole from a massive companion, HDE~226868, most likely via a focused wind. The value of the mass of the black hole is subject to controversy, it is in the range of $M\simeq (5$--$15) \msun$ according to Herrero et al.\ (1995) or $M\simeq (14$--$27) \msun$ (Zi{\'o}{\l}kowski 2005; see also Gies \& Bolton 1986). In this paper we will assume $M=15 \msun$. The distance, $D$, is most likely within $D\simeq 2.1\pm 0.2$ kpc (Zi{\'o}{\l}kowski 2005 and references therein); hereafter we adopt $D=2$ kpc.

As other black hole binaries Cyg X-1 presents various X-ray spectral states. The main ones are the so-called  Low Hard State (hereafter, LHS) and the High Soft State (hereafter HSS, see Done, Gierli{\'n}ski \& Kubota 2007 for a comprehensive review). 
In the LHS, the  isotropic bolometric luminosity $L_{\rm h, obs}$ is rather stable and close to 2$\times$10$^{37}$ erg s$^{-1}$ (see e.g. Gierli{\'n}ski et al. 1997;  Frontera et al. 2001;  Di Salvo et al. 2001, McConnell et al. 2002; Cadolle Bel et al. 2006).

Cyg X-1 is also a variable radio source. Multi-wavelength observations in the LHS have shown the presence of a flat-spectrum radio emission which  is produced by synchrotron emission from relativistic electrons in a compact, self-absorbed jet (Hjellming \& Johnston 1988).  This  jet was later resolved at the milliarsec scale by Stirling et al. (2001). The average radio flux is about 15 mJy in the LHS. In the HSS this  radio emission is strongly suppressed, { indicating that the radio jets may not be produced}. {  Deep radio observations of the field of Cyg X-1 resulted in the discovery of a shell-like structure
which is aligned with the resolved radio jet (Gallo et al. 2005). This large-scale (5 pc in diameter) structure appears to be inflated by the inner radio jet}. Gallo et al. (2005) estimate that in order to sustain the observed emission of the shell, the jet of Cyg X-1 has to carry a kinetic power that is comparable to the bolometric X-ray luminosity $L_{\rm h, obs}$ of the binary system. 
Then Russell et al. (2007)  refined this estimate using H${\alpha}$ and \mbox{[O\,{\sc iii}]} measurements of the jet-powered nebula. They estimate that the total kinetic power of the double sided jet  is $L_{\rm J}$=(0.9--3) $\times 10^{37}$ erg s$^{-1}$.
If we adopt $L_{h}$=2 $\times$ 10$^{37}$ erg $s^{-1}$ as the typical X-ray luminosity in the hard state then $j=L_{\rm J}/L_{\rm h}$ is in the range 0.45--1.5.

The bolometric  luminosity does not change dramatically during the state transition to the soft state { (see e.g.  Frontera et al. 2001;  Zdziarski et al. 2002; Malzac et al. 2006; Wilms et al.  2006)}. In the following we  combine this fact with the estimate of the jet power to constrain the initial and terminal jet velocities, as well as the nature of the X-ray emitting region. The structure of the paper is as follows. In section~\ref{sec:emae} we write down the equations for the energy budget of the accretion/ejection flow and show that the accretion efficiency cannot increase much during the state transitions { from LHS to HSS}. Then in section~\ref{sec:tjv} we show that under reasonable assumptions the terminal jet velocity must be relativistic. Then in section~\ref{sec:sizeandinitialvelocity} we show that if the terminal jet velocity is indeed relativistic and the Thomson depth along the radius of the base of the jet is larger than unity, then the initial velocity of the plasma in the base of the jet must be non-relativistic (and probably very low). These results and their caveats are then discussed in the context of current accretion models in section~\ref{sec:discussion}.

\section{Energetics of Cyg X-1}

\subsection{Energy and mass budget, accretion efficiencies}\label{sec:emae}

We define $\dot{M}_{\rm h}$ as the mass accretion rate in the hard state and at sufficiently large distances from  the black hole so that it is representative of the amount of material available for both accretion and ejection. A fraction $f_j$ of the accreting mass is ejected, the rest is swallowed by the black hole. 
The total  power output of the system (jet+radiation) must be equal to that extracted through accretion:
\begin{equation}
L_{\rm J}+L_{\rm h}=\eta_{\rm h} (1-f_j)\dot{M}_{\rm h}c^2,
\end{equation}
where $\eta_{\rm h}$ is the accretion efficiency in the hard state, and $c$ the speed of light. The accretion efficiency characterises the total energy available for both radiation and jet production. The radiative efficiency can be expressed as : 
\begin{equation}
\frac{L_{\rm h}}{\dot{M}_{\rm h}c^2}= \eta_{\rm h} \frac{1-f_j}{1+j} 
\label{eq:radef}
\end{equation}
Therefore if we knew both the radiative and total accretion efficiency, the known estimate of the jet power would allow us to evaluate the ejected mass fraction $f_{\rm j}$. In turn, the ejected mass fraction would yield the jet velocity.
However both efficiencies are poorly constrained, they depend on the unknown and possibly complex dynamics of the accretion flow in the hard state (possible role of advection, possible non-keplerian orbits).  The form of the accretion flow in the hard state is still a matter of debate. In order not to rely on any specific model we will consider that $\eta_{\rm h}$ may take any value comprised between 0 and 1.  

Then, in order to obtain some meaningful constraints we will have to combine this with the simpler situation that occurs in the HSS. 
Indeed, in the soft state there is observationally no evidence for a jet. We will therefore assume that the jet is not produced in the HSS, or at least that it is energetically negligible. This means that, in the soft state, the radiative efficiency and the total accretion efficiency are identical:
\begin{equation}
L_{s}=\eta_{\rm s}\dot{M}_{\rm s} c^2,
\label{eq:etasoft}
\end{equation}
where $L_{\rm s}$, $\eta_{\rm s}$ and $\dot{M}_{\rm s}$ are respectively  the luminosity, accretion efficiency and mass accretion rate in the HSS.
There is overwhelming evidence that accretion in the soft state  proceeds predominantly through a geometrically thin disc. The accretion efficiency of such a disc is theoretically limited in the range $\eta_{\rm s}=$0.06--0.4 depending on the black hole spin. 

Then, using, equations~(\ref{eq:radef}) and (\ref{eq:etasoft}), we define:
\begin{equation}
\lambda=\frac{L_{\rm s}}{L_{\rm h}}\frac{\dot{M}_{\rm h}}{\dot{M}_{\rm s}}=\frac{\eta_{\rm s}}{\eta_{\rm h}}\frac{1+j}{1-f_{\rm j}}.
\label{eq:lamb}
\end{equation}
$\lambda$ is nothing else than  the ratio of the HSS to LHS radiative efficiencies. The value of this ratio is a crucial parameter determining the energetics of the system. 
It is reasonably well constrained by the observations. 
In the HSS, the isotropic bolometric luminosity is somewhat higher than in the LHS.  Observations performed at different epochs with different instruments provide estimates of the isotropic bolometric luminosity in the HSS,  $L_{\rm s, obs}\simeq$ (6.2--7.2) $\times10^{37}$ erg~s$^{-1}$ (Gierli{\'n}ski et al. 1999; Frontera et al. 2001; McConnell et al. 2002). 
Therefore, during the state transition the observed luminosity jumps at most by a factor $L_{\rm s, obs}/L_{\rm h, obs}\simeq$ 3--4 (Zdziarski et al. 2002). 
We note that owing to possible anisotropy of the X-ray emission,  the observed isotropic luminosities may be different from the real intrinsic luminosities entering the definition of $\lambda$. However, as will be shown in section~\ref{sec:beam}, the effects of radiation beaming are expected to be weak and usually tend to reduce $\lambda$.
Moreover the transition to soft state is likely to be triggered by an increase in mass accretion rate and therefore $\dot{M}_{\rm h}/ \dot{M}_{\rm s}<1$. For these reasons, we can safely constrain $\lambda\leq4$. 

If, as it is widely believed, accretion proceeds in an advection dominated  accretion flow in the hard state (Narayan \& Yi 1994; Esin et al. 1997) then one would expect to have $\eta_{\rm s}>\eta_{\rm h}$.  In fact,  equation~(\ref{eq:lamb}) implies $\eta_{\rm s}/\eta_{\rm h}\leq\lambda/(1+j)$.
As we know that $\lambda\leq4$ and $j\geq0.45$, this gives $\eta_{\rm s}/\eta_{\rm h}\leq2.75$. Therefore, the accretion efficiency does not increase dramatically across the transition { from LHS to HSS}. Moreover, since $\eta_{\rm h}$ must be lower than unity, and $\eta_{s}$ is larger than 0.06,  we can set  very conservative limits on the efficiency ratio: $0.06\leq\eta_{\rm s}/\eta_{\rm h}\leq2.75$. 

In the following we will use  $\lambda=3$, $j=1$, $\eta_{\rm s}=\eta_{\rm h}=0.1$ as typical values. $\eta_{\rm s}$=0.1 corresponds to a black hole with a moderate spin parameter $a=0.4$. We note however that the recent spectroscopic results of Miller et al. (2009) suggest a nearly non spinning black hole ($a<0.06$) in which case the efficiency $\eta_{\rm s}$ would be close to 0.06.
We will also consider the unlikely combination $\lambda=4$, $\eta_{s}=0.06$, $\eta_{\rm h}=1$, $j$=0.45 as an extreme set of parameters used to provide a robust limit on the jet velocities.  

Finally,  we can write the fraction of ejected material as:
\begin{equation}
f_j=1-\frac{\eta_{\rm s}}{\eta_{\rm h}}\frac{1+j}{\lambda}
\label{eq:fj}
\end{equation}
For the fiducial model the fraction of ejected material is $f_{\rm j}=1/3$. For the extreme parameters most of the accreting material is ejected ($f_{\rm j}=0.978$).

\begin{figure}

\includegraphics[width=\columnwidth]{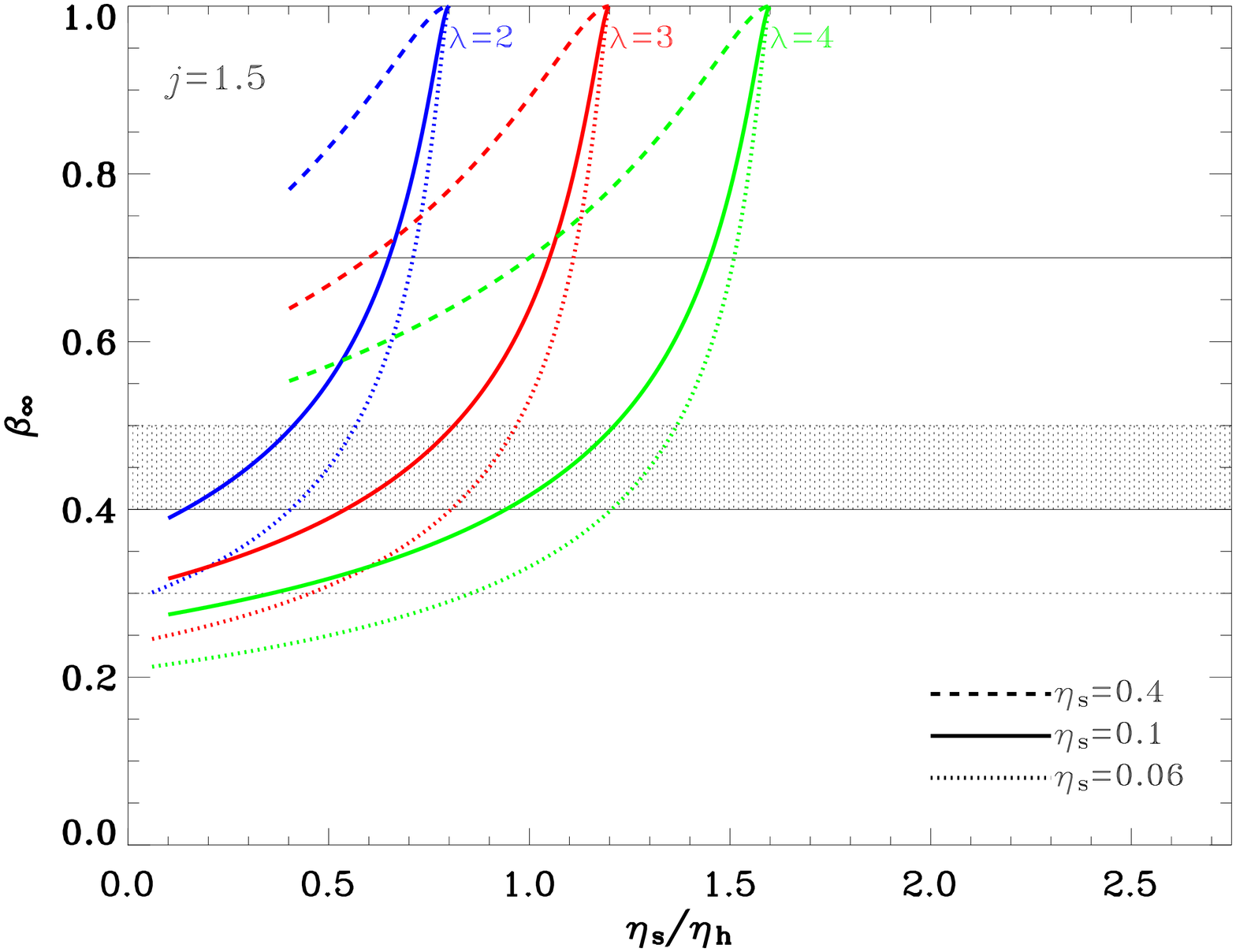}
\includegraphics[width=\columnwidth]{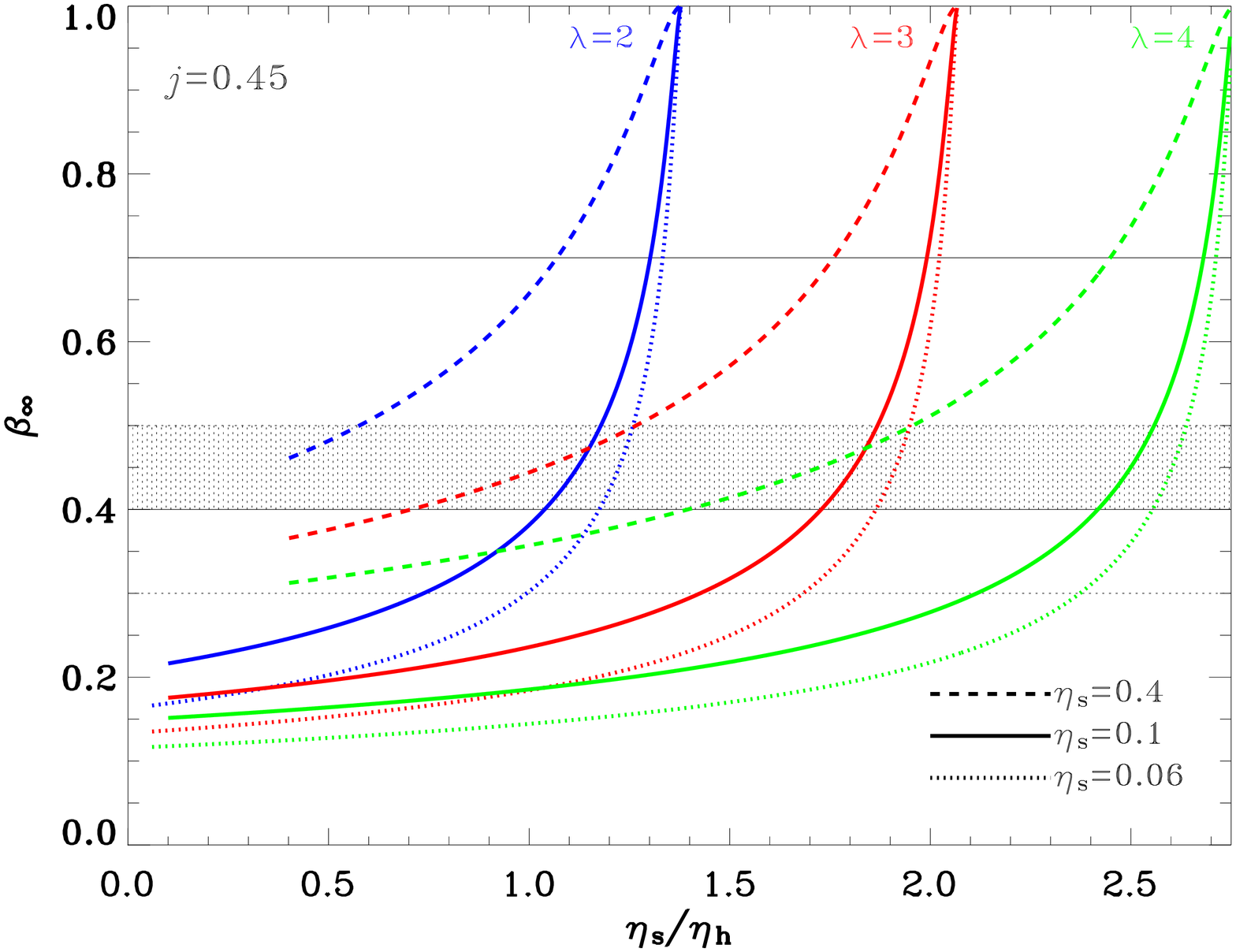}

\caption{ The jet terminal velocity as a function of of the accretion efficiency ratio $\eta_{\rm s}/\eta_{\rm h}$ for a jet power of 3$\times 10^{37}$ erg s$^{-1}$ (top panel) and 9 $\times 10^{36}$ erg s$^{-1}$ (bottom panel). In both panels the blue, red and green curves stand for $\lambda$=2,3 and 4 respectively. The dashed, full and dotted curves stand for $\eta_{\rm s}$=0.06, 0.1 and 0.4 respectively. The curves are plotted only in the range of $\eta_{\rm s}/\eta_{\rm h}$ for which there is a solution ($f_{\rm j}>0$) and $\eta_{\rm h}<1$. The black thin solid lines show the upper and lower limits provided by the radio imaging and radio X-ray correlations (Gleissner et al. 2004). The {black thin dotted lines} show the constraints from the modelling of the super-orbital periodicity (Ibragimov et al. 2007). The horizontal grey stripe shows the overlapping region between those two constraints. }
\label{fig:betainfdeet}
\end{figure}

\begin{figure}
\includegraphics[width=\columnwidth]{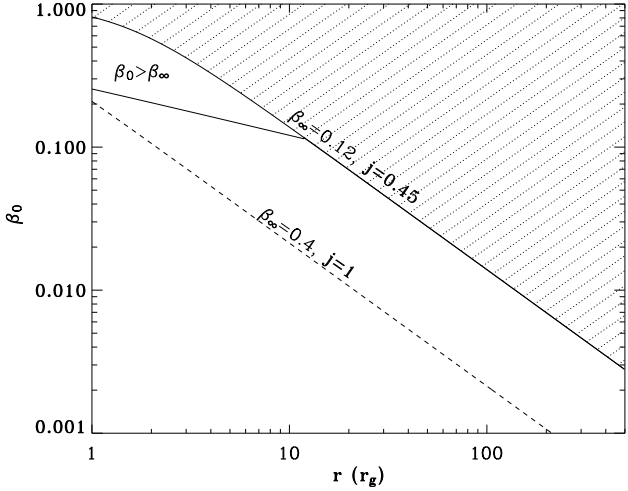}
\caption{Constraints on the jet initial velocity and radius in Cyg X-1. The hatched region shows the exclusion area delimited by the relation $\beta_0$ vs $r_0$  in the extreme case $\lambda=4$, $\eta_{\rm s}=0.06$, $\eta_{\rm h}=1$, $j=0.45$, corresponding to a minimal terminal velocity $\beta_{\infty}=0.12$.  The region above the thin continuous line is allowed in principle but requires the jet to slow down ($\beta_0>\beta_{\infty}$).  The dashed line shows the relation between $r$ and $\beta_0$ for $j=1$ a terminal jet velocity  $\beta_{\infty}=0.4$ which is favoured by the radio observations and also corresponds to our fiducial model with $\lambda=3$, $\eta_{\rm s}=\eta_{\rm h}=0.1$. In all curves the optical depth $\tau=1$ and the black hole mass is 15 $\msun$.}
\label{fig:betader}
\end{figure}

\subsection{Terminal jet velocity}\label{sec:tjv}

Based on the absence of detection of the  counter jet, Stirling et al. (2001) give a lower limit on the bulk velocity of the radio jet of $\beta_{\infty}>0.6$. Similar considerations and the lack of response of the radio emission on short time-scales led Gleissner et al. (2004) to constrain the jet velocity in the range $0.4<\beta_{\infty}<0.7$. Ibragimov, Zdziarski and Poutanen (2007) find a similar result $0.3<\beta_{\infty}<0.5$ from the modelling of the super-orbital modulation observed in the X-ray and radio band. In the following we compare these estimates with the one provided by the jet and disc energetics. 

The jet  kinetic power is:
\begin{equation}
L_{\rm J}=jL_{\rm h}=f_j\dot{M}_{\rm h} (\gamma_\infty-1)c^2,
\label{eq:jetpo}
\end{equation}
where $\gamma_{\infty}$ is the  'terminal' Lorentz factor of the jet at large distance from the black hole, in the region where most of its power is deposited  in the interstellar medium. Combining equations~(\ref{eq:jetpo}) and (\ref{eq:fj}),  it can expressed in terms of the accretion efficiencies:
 \begin{equation}
\gamma_\infty=1+\frac{\eta_{\rm s} j}{\lambda-(j+1)\frac{\eta_{\rm s}}{\eta_{\rm h}}}
\label{eq:ginf}
\end{equation}

For a typical accretion efficiency $\eta_{\rm s}=\eta_{\rm h}=0.1$, assuming $j=1$ and $\lambda=3$, we derive $\beta_{\infty}\simeq0.4$, in remarkable agreement with the constraints based on the radio observations.  Fig.~\ref{fig:betainfdej} explores how this result depends on the efficiency parameters. 

The red hatched area of Fig.~\ref{fig:betainfdej} shows the region of the $\beta_{\infty}$-$j$ plane that is allowed by the uncertainty on the accretion efficiency $\eta_{\rm s}$ (for $\lambda=3$ and $\eta_{\rm h}=\eta_{\rm s}$).  This region overlaps widely with the constraints from the radio observation and allows $\beta_{\infty}$=0.2-0.9.  

As mentioned above $\lambda$ could be lower than $3$, if so the velocity of the jet must be larger. As shown on Fig.~\ref{fig:betainfdej}, for $\lambda=2$, $\beta_{\infty}$ is in the range 0.3--1.  For $\lambda=2$ (and $\eta_{\rm s}=\eta_{\rm h}$) there is  no solution with  $j>1$ (see equation~(\ref{eq:lamb})).  

{ Fig.~\ref{fig:betainfdeet} shows that the terminal jet velocity increases with the ratio $\eta_{\rm s}/\eta_{\rm h}$ and may become very large.  Indeed for a given $j$ and $\lambda$ there is a critical value of the ratio $\eta_{\rm s}/\eta_{\rm h}$ for which the ejected fraction $f_{j}$ vanishes (see equation~\ref{eq:fj}) and the jet terminal velocity becomes infinite. There is no solution above this critical value of $\eta_{\rm s}/\eta_{\rm h}$.  As seen on Fig.~\ref{fig:betainfdeet}, for $j$=1.5 the maximum possible $\eta_{\rm s}/\eta_{\rm h}$ are 0.8, 1.2, 1.6 for $\lambda$=2,3 and 4 respectively.  For a weaker jet, the critical values of $\eta_{\rm s}/\eta_{\rm h}$ are increased. For $j=0.45$, they are  1.38,2.05 and 2.75 for $\lambda$=2,3 and 4 respectively. This confirms that the accretion efficiency cannot increase much during the transition from LHS to HSS.}

{ Overall the terminal bulk velocity increases with $\eta_{s}$, $j$, and $\eta_{\rm s}/\eta_{\rm h}$ and decreases with $\lambda$. }
The dependence of $\beta_{\infty}$ on $j$ in the extreme case $\eta_{\rm s}=0.06$, $\eta_{\rm h}=1$, $\lambda=4$ is shown in Fig~\ref{fig:betainfdej}. It provides a conservative lower limit on the terminal jet velocity $\beta_{\infty}> 0.12$.  
We conclude that, if the nebula of Cyg X-1 is powered by the jet kinetic power, the terminal velocity of the jet must be at least mildly relativistic. 
 
\subsection{Size and velocity of the X-ray emitting region} \label{sec:sizeandinitialvelocity}

Taking into account the fact that the jet is two-sided, mass is ejected at a rate:
\begin{equation}
\dot{M}_{\rm J} =2\pi R_0^2\gamma_0\beta_0c  n_0 m_p
\label{eq:mdotj2}
\end{equation}
where $m_p$ is the proton mass, $n_0$ is the comoving frame density at the base of the jet,  $R_0$ its radius, $\gamma_0$ and $\beta_0c$ are the Lorentz factor and velocity of the material entering the jet.  
Then, combining equations~(\ref{eq:jetpo}) and (\ref{eq:mdotj2}) and setting $L_{\rm h}$=2$\times 10^{37}$ erg s$^{-1}$, we find: 

\begin{equation}
r_0\gamma_0\beta_0=2 \times 10^{-2} \frac{j}{\tau(\gamma_{\infty}-1)}\frac{15 M_{\sun}}{M}
\label{eq:rp}
\end{equation}
where $\tau=n_0\sigma_T R_0$ is the Thomson depth along the radius of the base of the jet, $r_0=R_0/R_G$ and $R_G$ is the gravitational radius. Assuming there is no jet in the soft state and that the optical nebula is powered by the jet kinetic power,  equation~(\ref{eq:rp}) may be combined with equation~(\ref{eq:ginf}) to express the initial jet velocity in terms of the accretion efficiencies parameters $\eta_{\rm s}
$ and $\eta_{\rm h}$ and the observationally constrained parameter $\lambda$.

 For now, let us assume that the X-ray emission in the hard state is produced at the base of the jet. If so $\tau$ should correspond to the Thomson depth of the X-ray emitting medium.  Let us now further assume that the X-ray emission is produced by Comptonisation. { This standard assumption is overwhelmingly supported by the data (see e.g. Zdziarski  \& Gierli{\'n}ski 2004). Then, using Comptonisation models, the Thomson depth $\tau$ can be measured directly through spectral fits.}
 
{  There is  a continuum of spectral parameters between the hard and soft state rather than sharp transition. To define the spectral states one needs to set some boundaries on the spectral parameters. The exact value of these boundaries are somewhat arbitrary.  It is customary to consider that the source is in the hard state when the X-ray  photon index $\Gamma$ is in the range 1.5-2.1. But the 'softer' hard state spectra are less common and appear usually when the source is about to make a transition or failed transition (e.g., Malzac et al. 2006). These softer spectra  are called 'hard intermediate states' by some authors (Homan \& Belloni  2005; Del Santo et al. 2008). They  are probably not representative of the stable hard state spectra associated with steady compact radio jets. 
For this reason in this paper we will consider  the parameters when the source is far away from the transition i.e. when the spectrum is really hard $\Gamma\simeq1.6$. }
 
 { When the source is clearly in the hard state}, spectral fits with Comptonisation models yield a Thomson optical depth $\tau_T$ in the range 1--3 and electron temperature $kT_{\rm e}\simeq$ 60--100 keV (see e.g. Gierlinski et al. 1997; Frontera et al. 2001; Cadolle Bel et al. 2006)\footnote{An earlier spectral analysis using EXOSAT data suggested the optical depth could be lower $\tau_T\sim0.3$ and the temperature larger $kT_{\rm e} \sim150$ keV (Haardt et al. 1993). Nevertheless, more recent, better quality data confirm the optical depth is large. For $\tau_{\rm T}<1$ the individual Compton scattering orders become apparent in the high energy spectrum, producing bumps that are not observed. Also the higher temperature does not provide a good fit to the very sharp high energy cut-off.}
 depending on the details of the models and the observation. { It is also apparent in the fig. 6 of Ibragimov et al. (2005) and fig. 5 of Wilms et al. (2006) that the hardest spectra of Cyg X-1 are concentrated in a small region of the $\tau_{\rm T}$-$kT_{\rm e}$ plane with $\tau_{\rm T}\ge1$.}
 
 Comptonisation models used to fit the data assume a corona with a simple geometry such as a sphere or a slab. The Thomson optical depth  is defined along the smallest dimension of the corona (i.e. the radius for the sphere, or the height of the slab).  Only the smallest dimension is relevant since photons escape preferentially along this direction. In the case of emission in the base of the jet, the geometry will be cylindrical. If the scale height of the X-ray emitting region is larger than $R_0$ then we should have $\tau=\tau_T\simeq$1--3. If on the contrary the vertical extension of the corona is smaller than $R_0$ we simply have $\tau>$1--3. In any case, the optical depth  $\tau$ entering in equation~(\ref{eq:rp}) is larger than unity.
 
Then, as a consequence of the large optical depth of the base of the jet,  either the corona is tiny ($r<4r_g$) or the velocity of the corona is very small. 
Indeed, for  a given $r$, setting $\beta_{\infty}=0.12$ (i.e. to the minimum value allowed by the considerations of section~\ref{sec:tjv}) in equation~(\ref{eq:rp}) provides an upper limit on the initial jet velocity.  Fig.~\ref{fig:betader} shows that for $\tau=1$ this limit implies that for any reasonable jet section ($r_0\simeq 10-100$) the initial velocity must be non-relativistic ($\beta_{0}<$0.1). In the far more likely case of a jet terminal velocity $\beta_{\infty}=0.4$, that is consistent with both the previous radio estimates and the present ones, we find that $\beta_0$ must be  lower than a few $10^{-2}$ for any reasonable $r$.

We conclude that, if the Comptonising corona constitutes the base of the jet (or more generally if the base of the jet has a Thomson depth $\tau>1$) then its vertical bulk velocity is non-relativistic. 

\section{Beaming effects}\label{sec:beam}
\begin{figure}
\center
\includegraphics[width=\columnwidth]{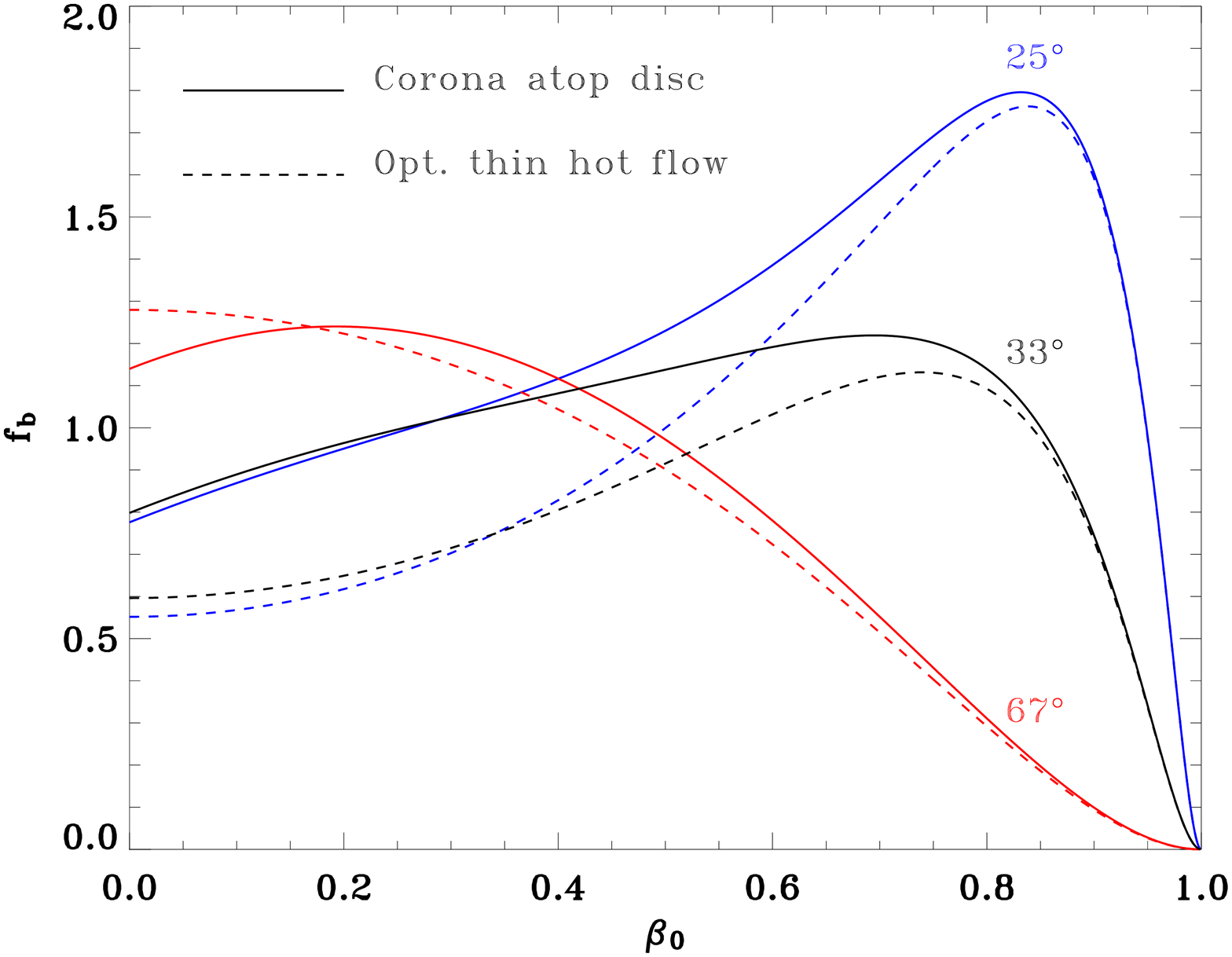}
\includegraphics[width=\columnwidth]{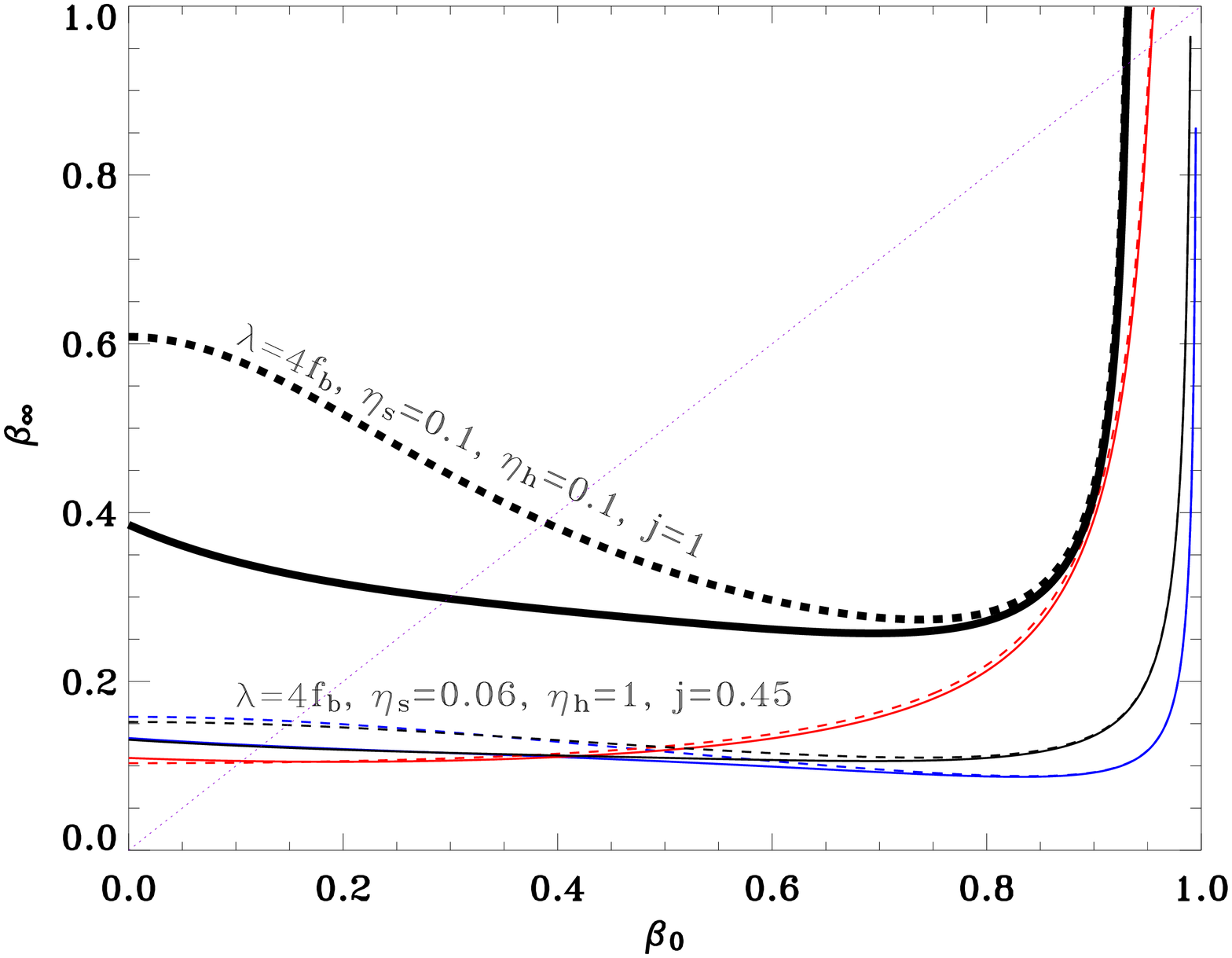}
\includegraphics[width=\columnwidth]{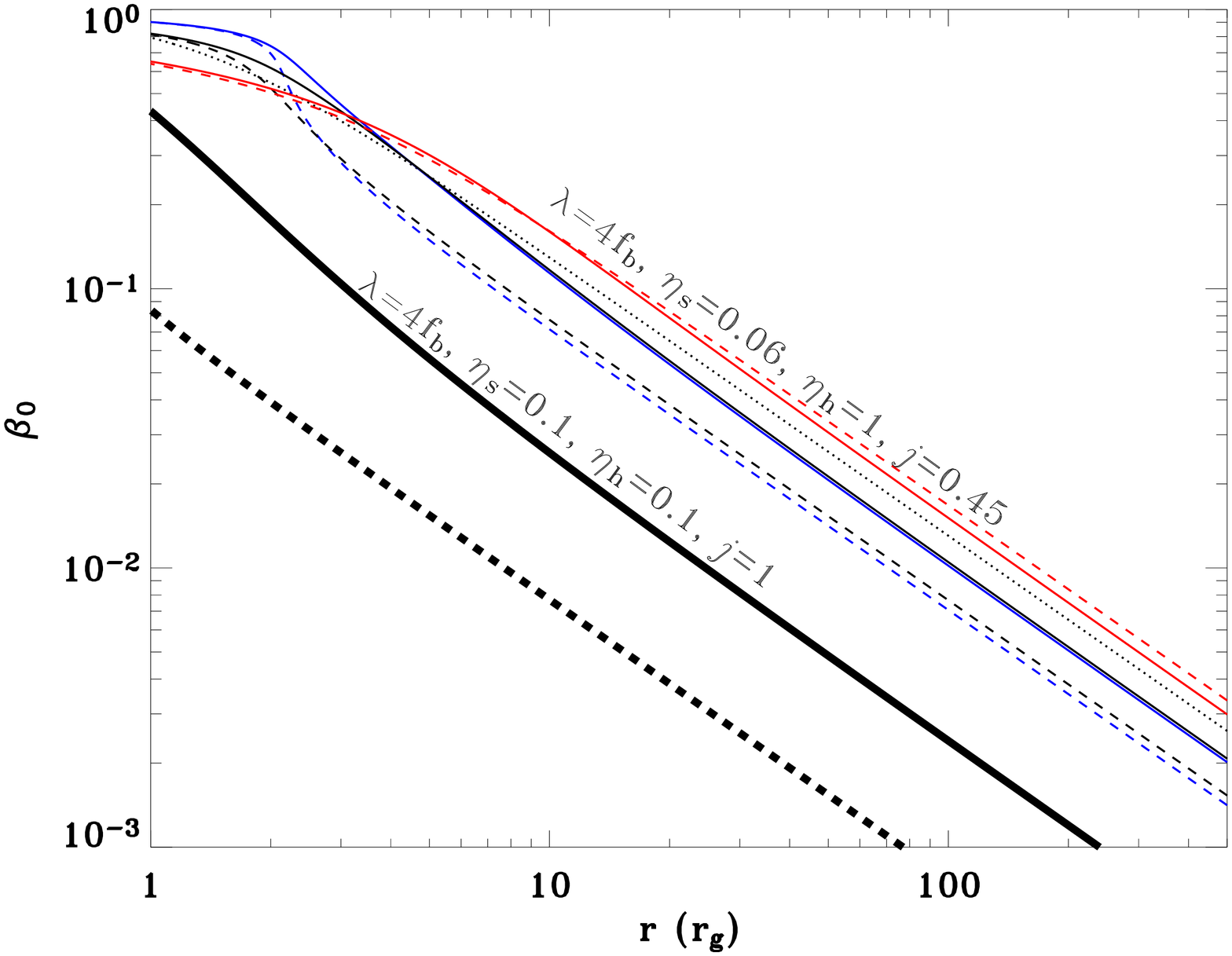}
\caption{Effects of beaming and anisotropy. Top panel: dependence of the beaming factor  $f_{\rm b}$ on the velocity of the X-ray emitting region $\beta_0$. Middle panel: effects on the minimum terminal velocity, $\beta_{\infty}$ as a function of $\beta_0$.  In the parameter space below the diagonal dotted line the terminal jet velocity is lower than the initial jet velocity. Bottom panel: effects of beaming on the maximum initial jet velocity. $\beta_{0}$ as a function of $r$. In all panels the blue curves stand for an inclination of 25 degrees, black for 33 degree, and red for 67 degree.
The solid lines assume a corona above an accretion disc (making use of equation~(\ref{eq:adc})) while the dashed lines dashed represent the results for a corona without optically thin disc and taking into account the radiation from the opposed side of the corona (using equation~(\ref{eq:transpa})).
The  thick black curves are for the fiducial model $\lambda=4f_{\rm b}$, $\eta_{\rm s}=\eta_{\rm h}=0.1$ and $j=1$. While the other thinner curves are for the limiting case $\lambda=4f_{\rm b}$, $\eta_{\rm s}=0.06$, $\eta_{\rm h}=1$ and $j=0.45$.} 
\label{fig:beam}
\end{figure}

In our estimates of the ratio of luminosities of the two spectral states, we have so far neglected the anisotropy of the X-ray emission. 
In the HSS  this anisotropy is due to a geometric effect: the emission is produced in the  thin disc plane. In the LHS, the X-ray emitting region has a velocity $\beta_0$ leading to Doppler beaming of the radiation.  This could  affect our estimates of $\lambda$ and the constraint on the jet velocities.  In order to take these effects into account  we rewrite $\lambda$ as:
\begin{equation}
\lambda=\frac{L_{\rm s, obs}}{L_{\rm h, obs}}\frac{\dot{M}_{\rm h}}{\dot{M}_{\rm s}}f_{\rm b},
\end{equation}
where $L_{\rm s, obs}$ and ${L_{\rm h, obs}}$ are the isotropic luminosities observed in the soft and hard state respectively.  $f_{\rm b}$ is a factor accounting for beaming effects and anisotropy. We know that the ratio  $\frac{L_{\rm s, obs}}{L_{\rm h, obs}}\simeq$(3--4)  and that $\frac{\dot{M}_{\rm h}}{\dot{M}_{\rm s}}\la 1$. 

We have now to estimate $f_{\rm b}$. Since in the soft state the luminosity is predominantly emitted as blackbody radiation, and since $L_{\rm s}$ is the power emitted by the two-sides of the disc, we have $L_{\rm s, obs}\simeq 2 L_{\rm s}\mu$ (with $\mu=\cos{i}$ and $i$ is the angle between the line of sight and the normal of the accretion disc).  Then following Beloborodov (1999) for the Doppler beaming effects of the X-ray emitting region with bulk velocity $\beta_0$, $f_{\rm b}$ can be evaluated as follows:
\begin{equation}
f_{\rm b}=\frac{1+\chi}{4\gamma_0^4\mu(1-\beta_0\mu)^3}.
\end{equation}
Because we consider the total energy radiated by both faces of the disc, this expression differs from the formula given by Beloborodov (1999) by a factor $1/2$. The factor $\chi$ accounts for  additional sources of radiation in the hard state (other than the direct emission from the corona). 
For instance, in the case of an accretion disc corona above a cold thin disc some level of disc reprocessing and reflection is expected. Then, assuming isotropy at the disc surface:
\begin{equation}
\chi\simeq 2\mu(1-\beta_0\mu)^3\frac{1+\beta_0/2}{1+\beta_0^2}.
\label{eq:adc}
\end{equation}

Alternatively, in the absence of a cold disc $\chi$ may account for the X-ray  radiation coming from the opposite side of the accretion flow and transmitted through the corona. Then in the optically thin limit:
\begin{equation} 
\chi\simeq\left(\frac{1-\beta_0\mu}{1+\beta_0\mu}\right)^3.
\label{eq:transpa}
\end{equation}

On the basis of analysis of absorption lines Gies \& Bolton (1986) estimated i=33$\pm$5 degrees. The polarimetric measurements of Dolan \& Tapia (1989) yield $i$ in the range 25--67 degrees. The upper panel of  Fig.~\ref{fig:beam} shows that in the allowed range of inclinations the $f_{\rm b}$ factor is of order of unity except for large $\beta_0$ when the Doppler beaming actually decreases significantly $f_{\rm b}$. Therefore the effects of anisotropy will tend to increase the value of $\beta_{\infty}$ rather than the opposite.  
For instance we can set  $\lambda=4f_{\rm b}$, $\eta_{\rm s}=0.06$, $\eta_{\rm h}=1$ and $j=0.45$ and compute the resulting minimum $\beta_{\infty}$ as a function of $\beta_0$. The middle panel of Fig.~\ref{fig:beam} shows that at small $\beta_0$ our lower limit is not significantly affected and stays in the range 0.1--0.2. If the initial velocity approaches the speed of light the terminal velocity must be large. We also plot the $\beta_{\infty}$-$\beta_0$ relation expected for the more likely set of parameters  $\eta_{\rm s}=0.1$, $\eta_{\rm h}=0.1$ and $j=1$, in this case the terminal jet velocity must be $>0.3c$, again in agreement with the constraints from the radio data. 

As shown in section~\ref{sec:sizeandinitialvelocity},  a lower limit on the terminal jet velocity sets an upper limit on the initial velocity at the base.  The bottom panel of Fig.~\ref{fig:beam} shows how the constraints discussed in section~\ref{sec:sizeandinitialvelocity} may  be affected by beaming.
Keeping  $\lambda=4f_{\rm b}$, $\eta_{\rm s}=0.06$, $\eta_{\rm h}=1$ and $j=0.45$,  the upper limit on $\beta_0$ as a function of $r$ is not changed qualitatively.  Moreover except for the largest possible inclination angle, beaming tends to reduce $\beta_0$. For the most likely inclination of 33 degree, the upper limit on $\beta_0$ is reduced by a factor of $\simeq 2$.
For the fiducial set of parameter, $\eta_{\rm s}=\eta_{\rm h}=0.1$ $j=1$  and an inclination of 33 degree,  $\beta_0<0.01$ for any reasonable $r$.

\section{Discussion}\label{sec:discussion}

\subsection{Consequences for the jet structure}

\subsubsection{Dark jet component, bending and misalignment}

Assuming that the jet is quenched in the soft state, and that the optical nebula around  Cyg X-1 is powered by the jet kinetic energy, we have set a  lower limit on the terminal jet velocity in the hard state of $\beta_{\infty}>0.1$. The actual velocity is most likely in the approximate range 0.3--0.8.  In section~\ref{sec:beam}, we investigated the possible effects of beaming of the X-ray radiation and found that they do not affect qualitatively these conclusions. 

Our estimates of $\beta_{\infty}$ are  the first constraints on the jet velocity obtained independently of any the radio measurements. This is important because  Gleissner et al. (2004) and Ibragimov et al.  (2007)  constrain only the velocity of the radio emitting material. There could be additional jet components which do not produce any radiation.  Indeed,  modelling the extended  radio emission observed by Stirling et al. (2001), Heinz (2006) finds that the kinetic power of the radio jet is several orders of magnitude lower than the estimates obtained from the observations of the optical nebula. In order to explain this puzzling result, Heinz (2006) suggests several alternatives. Among these he discusses the possibility that  the source of the kinetic energy powering the nebula is not the steady radio-emitting jet but some other dark component. Our results indicate that if there is a dark component its velocity is comparable to that of the radio emitting material. 

Modelling the radio orbital periodicity in terms of free-free absorption in the wind of the companion, Szostek \& Zdziarski (2007) also infer the presence of a second jet component, which must be not only dark but also slow ($\beta_{\infty}\sim5\times 10^{-2} c$).  
Indeed, in their model, the observed  phase lags between radio bands can be explained if the jet is bent. As shown by these authors, the bending may be caused by  a slow dark component. This slow component could  transport most of the material and energy. Our results are not consistent with this picture and  require an average velocity which is definitely relativistic.  
Perhaps, the jet is bent for a different reason, such as the effects of ram pressure from the wind of the companion star.

We also note that even if the jet is not bent, similar phase lags can be produced by free-free absorption (as in the model of Szostek \& Zdziarski 2007), provided that the jet average direction is not exactly perpendicular to the orbital plane but tilted by a few degrees. 
{  This would not be uncommon among X-ray binaries.  Indeed, Maccarone (2002) reports that in at least two microquasars (GRO J 1655-40 and SAX J 1819-2525) the observed relativistic jets appear misaligned. This is believed to  be caused by the spin of the black hole not being perpendicular to the orbital plane.  Then the central parts of the accretion disc are forced to rotate in the same plane as the black hole (Bardeen \& Petterson 1975).  The jets are produced by the inner part of the disc, perpendicular to it and therefore misaligned. But then the disc also exerts  a torque on the black hole which will eventually align the black hole spin (see e.g. Natarajan \& Pringle 1998; King et al. 2005). However the current theoretical estimates indicate that the alignement time-scales are likely to be at least a substantial fraction of the lifetime of these systems (Maccarone 2002). For Cyg X-1, a rough estimate can be obtained from equation 2.16  of  Natarajan \& Pringle (1998) which gives  an alignment time of order of 4$\times$10$^5$ yr (for a spin parameter $a=$0.06).  Unfortunately, the time since the formation of the black hole is poorly constrained. It could be comparable to the optical nebula lifetime,  i.e. (2--6)$\times$10$^4$yr   according to Russell et al. (2007), possibly much longer, but shorter than the time since the formation of the progenitor, i.e.  (3.5--6.5)$\times$10$^6$~yr according to stellar evolution models (Mirabel \& Rodrigues 2003). If the black hole was formed recently with a misaligned spin, the jet of Cyg X-1 could still be misaligned nowadays. This may be the cause of the radio lags.}

\subsubsection{Jet acceleration and mass loading}\label{sec:initvel}

{In section~\ref{sec:sizeandinitialvelocity} it was shown that if the Comptonising corona is being ejected to infinity  (or more generally if the base of the jet has a Thomson depth $\tau>1$) then its vertical bulk velocity is non-relativistic.  In the case of a Compton corona, the observation requires a Thomson depth larger than unity along the vertical scale height of the corona. Therefore, the velocity must be low over a vertical scale $h$ which is at least of order of $r/\tau$. Otherwise,  if acceleration occurred at a lower scale height, the corona would be depleted of its material and be optically thin.  Then, the jet material leaving this launching region has  to be gradually accelerated at larger distances in order to reach the mildly relativistic speed observed at large scales.

 It is not clear however which acceleration mechanism could produce such a velocity profile.  In magnetically driven jet models most of the acceleration occurs relatively close to the accretion disc (see e.g. Ferreira 1997; Casse \& Ferreira 2001).  In principle, more distant acceleration could be achieved by converting the gas internal energy into bulk kinetic energy through longitudinal pressure gradients. If most of the initial thermal energy can be converted into bulk kinetic energy, and neglecting other forms of internal energy,   the terminal velocity $\beta_{\infty}$ is of order of $\sqrt{\frac{6kT_{\rm i}}{m_{\rm p} c^2}}$, where $T_{\rm i}$ is the gas temperature at the base of the jet.  In consequence accelerating the jet up to mildly relativistic  velocities requires a temperature that is a few times virial:
$\beta_{\infty}>0.3 $ would imply $T_{\rm i}\ga2\times 10^{11}$ K. Such high temperatures are expected in two temperature accretion flows (Narayan \& Yi 1994). Similarly, Merloni \& Fabian (2001) elaborate on the possibility of launching strong thermally driven outflows from 2-temperature accretion disc coronae. We note however that in the case of Cygnus X-1, the ion temperature of the hot corona is likely to be too low. Indeed, Malzac \& Belmont (2009) show that the relatively large Thomson depth implies $T_{\rm i} \la 2 \times 10^{10}$ K  which appears too cold to drive a strong thermal outflow.
 
 Perhaps a more likely situation would be that  only a small fraction of the X-ray emitting material is loaded onto the jet and ejected to infinity. This would allow a Thomson depth  larger than unity in the corona and arbitrarily low in the base of the jet.  The  initial velocity could then be relativistic, depending on the fraction of ejected material. This would also avoid the requirement for most of the jet acceleration being produced at large distances from the black hole. 
This scenario would also be consistent with recent numerical simulations (e.g. Hirose  et al. 2004; McKinney 2006; Fragile \& Meier 2009) suggesting that the jet launching region is compact, initial launching velocity is relativistic,  with low mass loading onto the jets.}

\subsection{Consequences for accretion/ejection models}\label{sec:consequences}

\subsubsection{X-ray jet models}

When the importance of the connection between radio and X-ray emission was realised, it was proposed by several authors that the hard X-ray emission of the LHS could originate from the jet. Most of these models require or assume an initial velocity that is too large. In order to reproduce the observed X-ray luminosity, the external Compton model of Georganopoulos, Aharonian \& Kirk (2002) requires a jet power of $\sim 5 \times 10^{38}$ ergs which is immediately excluded by the estimates of Russell et al. (2007).
In the model by Reig, Kylafis  Giannios 2003 (see also Giannios, Kylafis \& Psaltis 2004; Kylafis et al. 2008) the soft photons from the accretion disc are upscattered by the jet.  In this model the optical depth is in the range $\tau=1.2-15$, and the radius of the base $r_0=75-300$, with the velocity $\beta_0\simeq0.5-0.8$. This is in clear breach of the constraints of section~\ref{sec:sizeandinitialvelocity}.

These conclusions do not apply to the jet model of Markoff et al.  (2001, 2005 herafter M05). In this model the optical depth at the base of the jet is very small ($\tau\ll1$) and makes it energetically possible to have a mildly relativistic initial jet velocity.
Indeed, M05 fit the spectral energy distribution of Cyg X-1 with a jet dominated model which attributes the X-ray emission to thermal synchrotron self-Compton in the base of the jet. These fits result in a jet power that is comparable to the X-ray luminosity and therefore consistent with the estimates of Gallo et al. (2005) and Russell et al. (2007). M05 do not quote explicitly the resulting optical depth at the base of the jet but it can be easily estimated from the other parameters:  they infer a radius at the base $r_0$ which is in the range the 4--10 (depending on the data set). The initial jet velocity is $\gamma_0\beta_0\simeq0.4$ and the terminal jet velocity $\gamma_{\infty}\simeq2$--3. Using equation~(\ref{eq:rp}) this implies an optical depth $\tau$ in the range 10$^{-3}$-- 3$\times10^{-2}$. This is much lower than what is usually found when the same data are fit  with thermal comptonisation models (i.e. $\tau>1$). In fact, in this model the optical depth is so small that the X-ray spectrum must be  produced through single Compton scattering. This requires very energetic comptonising leptons. In order to fit the RXTE data, the model requires an electron temperature $\simeq$ 3 MeV.
 
 We note however that such a combination of small size, very low optical depth and large temperature is physically impossible.  Indeed, as discussed below in section~\ref{sec:caveats}, in  Cyg X-1, the large luminosity and small emitting region make the compactness larger than unity.  Therefore  achieving a very small  optical depth and a temperature  $kT>m_{\rm e}c^2$ is impossible due to the effects of pair production. Svensson et al. (1984) studied the equilibrium properties of a thermal pair plasma. His Fig.~6 shows that for a compactness of order of a few (like in Cyg X-1), the optical depth must be at least 10 times larger than in this jet model.  We also made some simulations using the code of Belmont, Malzac \& Marcowith (2008). Setting the  optical depth of ionization electrons $\tau_p=0.01$  and a bolometric luminosity of 2$\times10^{37}$ erg s$^{-1}$ emitted in a region of 30 $R_{\rm G}$, we found a total  equilibrium optical depth $\tau_T\simeq0.6$ and a lepton temperature $kT_{\rm e}\simeq 300$ keV.  The exact values of the equilibrium temperature and optical depth depend on the strength of magnetic field that, following M05, we assumed close to equipartition with the radiating electrons.  The resulting X-ray spectrum was clearly different from both the observed  and the ones computed by M05 neglecting pair production. 
 
 Then we explored extensively the parameter space and found  that  it is not possible to have simultaneously a low optical depth and a temperature as large as the one required by the M05 model unless the size of the emitting region $r_0$ is larger  than $\sim 10^4$ $R_G$. On the other hand if the emitting region is very compact $r_0<10$ the pair optical depth can reach unity, in agreement with usual Comptonisation fits, however the equilibrium pair temperature is then too large ($>150$ keV) to reproduce the sharp cut-off that is observed in the hard X-ray spectrum around 100 keV.  
 
{ The parameters of the model of M05 are therefore inconsistent with the constraints from pair equilibrium.} Once the effects of pair production are taken into account, it seems impossible to fit the high energy spectrum of Cyg X-1 with this model.  
Finally we note that X-ray dominated jet  models also appear to be ruled out by the comparison of the properties of black holes and neutron stars (Maccarone 2005).

\subsubsection{Hot flow models}

Our results are consistent with the popular idea that the X-ray emission is produced in some sort of  hot accretion flow (e.g. Shapiro Lightman \& Eardley 1976; Narayan \& Yi 1996).
As we have shown, if the hot flow constituted the base of the jet (i.e.  was being ejected to infinity), its ejection velocity would have to be very  slow. Then, the  ejection time could be  comparable to,  or even longer than the viscous time in the inner part of the accretion flow. If so the corona would be accreted  before being ejected. It is therefore unlikely that a large fraction of the accreting material is ejected. 
Also for the reasons discussed in section~\ref{sec:initvel}, it is much more plausible that only a small fraction of the hot flow is loaded into the jet and ejected to infinity. 
In this framework, our considerations would be consistent with the coupled ADAF/jet model of Yuan, Cui \& Narayan (2005). We stress however that the accretion efficiency changes at most by a factor of $\simeq2.7$ during the state transition (see section~\ref{sec:emae}) and therefore Cyg X-1 cannot be strongly advection dominated in the hard state. In this context, radiatively efficient hot flow solutions such as the
luminous hot accretion flow model of Yuan (2001) would be favoured.
We also stress that in the X-ray emitting region of Cyg X-1, the proton temperature seems much lower than the predictions of standard ADAF models (Malzac \& Belmont 2009).  

\subsubsection{Dynamic accretion disc corona model}

Beloborodov (1999) and then Malzac, Beloborodov \& Poutanen (2001) showed that the hard X-ray emission of Cyg X-1 can be understood in terms of an accretion disc corona outflowing  with a midly relativistic velocity ($\beta\simeq 0.3$) atop a geometrically thin, optically thick accretion disc. Since this velocity is comparable to the infered terminal radio jet velocity and also comparable to the initial launching velocity of some popular jet models, it is tempting to associate this dynamic corona with the base of the jet. 
However, the typical parameters of a dynamic corona ($\beta\simeq0.3$) lie at the edge of the allowed parameter regime (see Figs~\ref{fig:betader} and \ref{fig:beam}). Besides the extreme accretion efficiencies required, this would  imply a very small jet section ($<$4 $r_g$) and may require the jet to decelerate so that $\beta_{\infty}<\beta_{0}$.  A far more likely situation would be that the whole corona is indeed accelerated to relativistic speed (as in the dynamic corona model)  but  only a small fraction of its material actually escapes to infinity. This could be related to the  velocity of the gas being lower than the escape velocity close to the black hole (see e.g. Ghisellini, Haardt \& Matt 2004). Indeed, within 20 $R_g$, the escape velocity ($\beta_{\rm esc}=\sqrt{2/r}$) is larger than 0.3$c$ and most of the ejected gas would be accreted again.  Whatever the origin of the X-ray emission (hot accretion flow or accretion disc corona) we favour a situation in which only a small fraction of the X-ray emitting material is loaded onto the jet and ejected to infinity.

\subsection{Caveats}\label{sec:caveats}

Let us now consider the possible limitations to our results.
First, our estimates of the jet terminal velocity and in particular the lower limit obtained for this velocity relies on our assumption that the optical nebula of Cyg X-1 is powered by the kinetic power of the jet. { We note that the possibility that the nebula is actually a background supernova remnant cannot be fully ruled out (Russell et al. 2007). Even if the} nebula is powered by the jet, it is possible that the energy of the jet is not dominated by kinetic energy (as in the case of a Poynting flux dominated jet for instance). If so, our analysis remains valid provided the $j$ parameter is reduced accordingly. Then, the estimated jet velocity would be lower. 
 
Another possible limitation is that our estimate of $\beta_{\infty}$ relies on the absence of jet in the soft state. In fact, we cannot exclude that the HSS may produce a jet of very high Lorentz factor. Such a jet could transport away a large fraction of the energy and yet be unobservable because our line of sight falls outside the beaming cone (Maccarone 2005) or because of the absence of a shocked deceleration region. In our framework, this would be equivalent to reducing the radiative efficiency in the HSS and we would then infer lower jet velocities. 
 
 However the good agreement with the estimates of the jet velocity obtained from the radio observations indicates that our assumptions are at least roughly valid. This suggests that  the jet is not strongly dominated by Poynting flux and that any jet formed in the HSS does not take away a substantial fraction of the energy.
We also note that even if this was the case, our constraints on the initial jet velocity would remain valid provided the radio estimates of the jet bulk velocity are correct. 

Because the electron to proton mass ratio is so low,  we have neglected the contribution of leptons to the jet kinetic power. Taking into account the leptons would amount to divide our estimates of $\gamma_{\infty}-1$  by a factor $1+\xi m_{\rm e}/m_{\rm p}$, where $\xi$ is the number of leptons per proton in the jet. We see that our results would be affected only if the jet composition is strongly dominated by electrons-positron pairs. Again, the agreement with the radio estimates for our typical set of parameters suggests  this is not the case. 

We note however that  an important contribution of  electron-positron pairs to the optical depth $\tau$ at the base of the jet would  significantly affect  the product $\gamma_0 \beta_0 r_0$ (see equation~(\ref{eq:rp})) which would be increased by a factor $\xi/(1+\xi m_{\rm e}/m_{\rm p})$. If pairs are important at the base, the initial velocity may be large. For a typical size of the X-ray emission of $r_0=30 R_g$ and $L_{h}=2\times10^{37}$ the compactness parameter is larger than unity ($l=8$) and pair production may be important. In order to investigate this possibility, we performed numerical simulations using the code {\sc eqpair} (Coppi 1992).  For the parameters producing a hard X ray spectrum similar to that of Cyg X-1, we found that $1\leq\xi\leq2$, as long as the size of the X-ray emitting region $r_0>10$ (i.e. $l<24$) and $\tau>1$. Therefore  if $r_0>10$  the effects of pairs would be weak, increasing the initial velocity by less than a factor of 2. However if the X-ray emitting region is smaller,  electron-positron pairs may dominate.  If so the initial velocity could indeed be relativistic. 

Finally, in our calculations, we have assumed that the source of jet power was in the accretion disc. We have ignored the possibility that the jet may tap a significant fraction of its energy from the black hole spin (Blandford \& Znajek 1977).  If this additional source of energy is important, the net effect is to increase the apparent efficiency $\eta_{\rm h}$ in the LHS. Our results would be affected only if this process can lead to $\eta_{\rm h}>1$.  We believe this is unlikely, specially if  the black hole spin in Cyg X-1 is as small as inferred by Miller et al. (2009).

\section{Conclusion}

The jet power estimates of Russell et al. (2007) and the fact that the luminosity of the X-ray source does not increase dramatically during the state transition  put some interesting constraints on the energetics of the accretion flow in Cyg X-1. Notably, the accretion efficiency cannot increase dramatically across the state transition { from LHS to HSS}. The jet bulk velocity must be relativistic ($\beta_\infty>0.1$) and, depending on the accretion efficiency, it  is likely to be in the range 0.3--0.8. Then if the Thomson optical depth at the base of the jet is larger than unity the initial jet velocity must be very low.  This is in contradiction with several jet models in which the X-ray emission is produced in the jet or its base. We also pointed out that the specific jet model of M05 appears inconsistent with the observed X-ray spectrum of Cyg X-1 once the effects of pair production are taken into account. Finally, both hot accretion flow and outflowing accretion disc corona models remain consistent with our results provided that only a small fraction of the X-ray emitting material is loaded on the jet. The X-ray emitting region and the jet therefore appear to be distinct although physically connected.

\section*{Acknowledgments}
This work is supported by CNRS and ANR. We thank Anna Szostek, Pierre-Olivier Petrucci and Andrzej Zdziarski for useful comments on the manuscript. 
JM thanks the Institute of Astronomy of Cambridge for hospitality.

\bsp

\label{lastpage}

\end{document}